\documentclass[pra,aps,a4paper,twocolumn,repreprint, superscriptaddress, longbibliography]{revtex4-2}

\usepackage{amssymb}
\usepackage{amsmath}
\usepackage{amsfonts}
\usepackage{graphicx}
\usepackage{color, soul}
\usepackage{natbib}
\usepackage{hyperref}
\usepackage{adjustbox}
\usepackage{tabularx}
\usepackage{breqn}
 \usepackage{multirow}

\begin{document}

	\title{Magnetoconductance and photoresponse properties of disordered NbTiN films}

	\author{M. Sidorova} 
	\affiliation{Humboldt-Universität zu Berlin$,$ Department of Physics$,$ Newtonstr. 15$,$ 12489 Berlin$,$ Germany}
	\affiliation{German Aerospace Center (DLR)$,$ Institute of Optical Sensor Systems$,$ Rutherfordstr. 2$,$ 12489 Berlin$,$ Germany}
	
	\author{A.D. Semenov}
	\affiliation{German Aerospace Center (DLR)$,$ Institute of Optical Sensor Systems$,$ Rutherfordstr. 2$,$ 12489 Berlin$,$ Germany}
	
	\author{H.-W. Hübers}
	\affiliation{Humboldt-Universität zu Berlin$,$ Department of Physics$,$ Newtonstr. 15$,$ 12489 Berlin$,$ Germany}
	\affiliation{German Aerospace Center (DLR)$,$ Institute of Optical Sensor Systems$,$ Rutherfordstr. 2$,$ 12489 Berlin$,$ Germany}
	
	\author{S. Gyger}
	\affiliation{Department of Applied Physics$,$ KTH Royal Institute of Technology$,$ SE-106 91 Stockholm$,$ Sweden}
	
	\author{S. Steinhauer}
	\affiliation{Department of Applied Physics$,$ KTH Royal Institute of Technology$,$ SE-106 91 Stockholm$,$ Sweden}
	
	\author{X. Zhang}
	\affiliation{State Key Laboratory of Functional Materials for Informatics$,$ Shanghai
		Institute of Microsystem and Information Technology$,$ Chinese Academy of
		Sciences (CAS)$,$ Shanghai 200050$,$ China}
	\affiliation{CAS Center for Excellence in Superconducting Electronics$,$ Shanghai
		200050$,$ China}
	
	\author{A. Schilling}
	\affiliation{Physics Institute$,$ University of Zürich$,$ Winterthurerstrasse 190$,$ 8057 Zürich$,$ Switzerland}

	\begin{abstract}
	We report on the study of phonon properties and electron-phonon coupling in thin NbTiN films, which are intensively exploited in superconducting devices. Studied NbTiN films with thicknesses less than 10~nm are disordered with respect to electron transport, the Ioffe-Regel parameter of $k_F l_e = 2.5-3.0$ ($k_F$ is the Fermi wave vector and $l_e$ is the electron mean free path), and inelastic electron-phonon interaction, the product $q_T l_e \ll 1$ ($q_T$ is the wave vector of a thermal phonon). By means of magnetoconductance and photoresponse techniques, we derived the inelastic electron-phonon scattering rate $1/\tau_{e-ph}$ and determined sound velocities and phonon heat capacities. In the temperature range from 12 to 20~K, the scattering rate varies with temperature as $1/\tau_{e-ph}\propto T^{3.45\pm0.05}$; its value extrapolated to 10~K amounts to approximately 16~ps. Making a comparative analysis of our films and other films used in superconducting devices, such as polycrystalline granular NbN and amorphous WSi, we found a systematic reduction of the sound velocity in all these films by about 50\% as compared to the corresponding bulk crystalline materials. A corresponding increase in the phonon heat capacities in all these films is, however, less than the Debye model predicts. We attribute these findings to reduced film dimensionality and film morphology. 
	\end{abstract}

\date{\today}
\maketitle

\section{Introduction}
\label{sec: Introduction}

Thin disordered NbTiN films are intensively exploited in various superconducting devices. They have become a prominent material for the fabrication of scientific and commercial superconducting nanowire single-photon detectors (SNSPDs) \cite{esmaeil2021superconducting} and already a decade ago were considered as an alternative to NbN films in hot-electron bolometers (HEBs) \cite{shurakov2015superconducting}. Both electron transport properties and phonon-related properties of these films  are of high importance because they determine the performance metrics of practical devices. For instance, the strength of electron-phonon coupling and the sound velocity along with the heat capacity of phonons jointly control the relative strength of electron heating and the cooling rate of electrons  and, therefore, determine the intermediate-frequency (IF) bandwidth in HEBs \cite{semenov2002hot, klapwijk2017engineering} or impact the timing jitter in SNSPDs \cite{vodolazov2019minimal}. Although in the literature, it was repeatedly indicated that NbN-based HEBs demonstrate larger IF bandwidths than NbTiN-based HEBs (see e.g. the review \cite{shurakov2015superconducting} and references therein), the explicit reason for this experimental fact has never been clarified. A lack of both theoretical and empirical description of properties of NbTiN and other films used in superconducting devices is often caused by their complexity, i.e. high level of disorder and reduced dimensionality with respect to different physical phenomena. NbTiN is a compound of two well studied nitrides, NbN and TiN, with very different electron-phonon (\textit{e-ph}) scattering times ($\tau_{e-ph}=1.8$~ns for TiN \cite{kardakova2013electron} and $\sim10$~ps for NbN \cite{sidorova2020electron} at 10~K). From this perspective, it is interesting to understand whether the $\tau_{e-ph}$ can be finely tuned by changing the Nb/Ti fraction $x$ in Nb$_x$Ti$_{1-x}$N. The matter is even more complicated since the superconducting properties of NbN films alone such as transition temperature, resistivity, and critical current density are strongly dependent on the film stoichiometry \cite{semenov2021superconducting}. It has been shown that SNSPDs based on amorphous superconductors, e.g. silicides such as WSi \cite{le2016high} or MoSi \cite{verma2015high}, can show favorable properties for the detection of low-energy infrared photons due to their lower superconducting transition temperature compared to Nb-based materials. The explicit role, if any, of the amorphous state in this improvement however remains unclear. 

In this study, we derive various properties of NbTiN films at low temperatures, in particular, the \textit{e-ph} scattering time, the sound velocity, and the phonon heat capacity, and carry out a comparative analysis with other films, NbN and WSi, intensively used in various superconducting devices. We show that, in all films, the sound velocity is reduced and the phonon heat capacity is increased as compared to the corresponding bulk crystalline materials. We suppose that this effect is exclusively controlled by modification of acoustic phonons due to reduced film dimensionality and film morphology.

\section{Experiment and results}
\label{sec: Section name}

NbTiN films with thicknesses 6 and 9~nm (samples $\#$L135 and $\#$L134, respectively, in Table~\ref{tab:transport_SC_parameters}) were deposited on 270~nm-thick SiO$_2$ layers  thermally grown on Si substrates. Details of the deposition process have been reported elsewhere \cite{steinhauer2020nbtin}. According to studies of films prepared under the same sputtering conditions \cite{zichi2019optimizing}, the compositions of our Nb$_x$Ti$_{1-x}$N films had a Nb/Ti fraction of approximately $x=0.6$ and polycrystalline granular structure with diameters of most grains in the range from 4 to 5~nm.

\subsection{Magneto-transport measurements}

\newcolumntype{c}{>{\centering\arraybackslash}X}
\begin{table*}[ht!]
	\centering
	\caption{Transport and superconducting parameters of studied here NbTiN films ($\#$L134, $\#$L135) and the parameters (taken from literature) of three other (NbN, WSi) superconducting films.  $RRR$ is the ratio between resistances at 300~K and at 25~K. The values of $k_F$ and $l_e$ are computed from the carrier density $n_e$ determined from Hall measurements at 25~K.}
	\label{tab:transport_SC_parameters}
	\begin{tabularx}{\textwidth}{@{}ccccccccccccccc@{}}	
		\hline \hline
		Material&Sample & $d$  & RRR  & $T_{C0}$ & $R_{SN}$      & $D$                      & $N(0)$                    & $B_{C2}(0)$ & $\xi(0)$ & $n_e$        & $k_F$  & $l_e$ & $k_F l_e$  \\
		& & (nm) &      & (K)      & ($\Omega$/sq) & (cm$^2$/s) & (J$^{-1}$m$^{-3}$) & (T)         & (nm)     & (m$^{-3}$) & (nm$^{-1}$) & (nm)  &        \\	
		\hline
		NbTiN&L135   													   & 6.0& 0.80 & 8.41 & 710.6    & 0.458 & 2.00$\times10^{47}$  & 13.9  & 4.9& 4.97$\times10^{28}$      & 11.4   & 0.22            & 2.5     \\
		NbTiN&L134  													   & 9.0& 0.83 & 9.51 & 381.4   & 0.472 & 2.40$\times10^{47}$ & 15.2  & 4.6& 5.39$\times10^{28}$    & 11.7   & 0.26             & 3.0   \\
		NbN\cite{sidorova2020electron} &2259     & 5.0& 0.79 & 10.74 & 529.5  & 0.474 & 3.11$\times10^{47}$ & 17.4   & 4.4& -   								        & -   &       -      & 1.2\footnotemark[1]   \\
		NbN\cite{sidorova2020electron} &A853   & 6.4& 0.71 & 8.35 & 954.0  & 0.339 & 1.89$\times10^{47}$ & 19.2   & 4.2& -   								        & -   &       -      & 0.9\footnotemark[1]   \\
		WSi\cite{zhang2016characteristics}& 4& 5.0& 0.96 & 4.08& 357.0  & 0.700  & 3.10$\times10^{47}$ & 6.7     & 7.0 & -   									   & -   &       -      & 1.8\footnotemark[1]   \\
		\hline	\hline
	\end{tabularx}
	\footnotetext[1]{For these films, the Ioffe-Regel parameter was computed as $k_F l_e = 3Dm_e/\hbar$ wth the free electron mass $m_e$.}
\end{table*}

\begin{figure}[h!]
	\centerline{\includegraphics[width=0.5\textwidth]{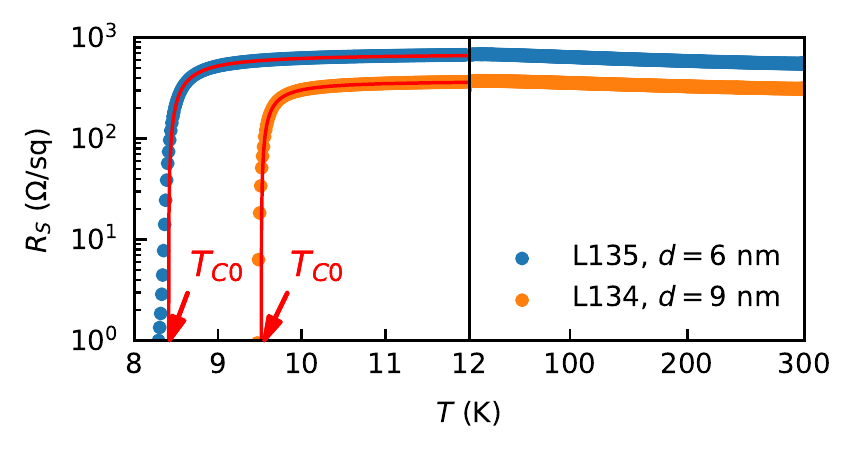}}
	\caption{Square resistance vs temperature at zero magnetic field (semi-logarithmic scale). Points: experimental data; curves: the best fits obtained with Eq.~\ref{eq:RT_2d} at temperatures, where $\ln(T/T_{C0}) \ll 1$, and extrapolated to higher temperatures.}
	\label{fig:RT}
\end{figure}

Resistance and Hall measurements were carried out by a standard four-probe technique in a Physical Property Measurement System (manufactured by Quantum Design) in magnetic fields applied perpendicular to the film surface. The van der Pauw method was used to determine the square resistance $R_S$ and eliminate the effect of the planar geometry for two-dimensional (2d) specimens. Fig.~\ref{fig:RT} shows the zero-field $R_S$ as a function of temperature. When temperature decreases from 300~K down to about 25~K, $R_S$ increases and reaches the maximum value of 705.9 and 383.8~$\Omega$/sq for the film $\#$L135 and $\#$L134, respectively, that is most likely due to Anderson localization. In the vicinity of superconducting transition, where the inequality $\ln(T/T_{C0}) \ll 1$ holds, we fitted the $R_S(T)$ data with the theory of fluctuation conductivity of Aslamazov and Larkin \cite{aslamasov1968influence} and Maki and Thompson \cite{maki1968critical, thompson1970microwave}. For 2d films, it is given by
\begin{equation}\label{eq:RT_2d}
	R_{S}(T) = \left[ \dfrac{1}{R_{SN}} + A_{2d}\dfrac{1}{16} \dfrac{e^2}{\hbar \ln(T/T_{C0})} \right]^{-1}.
\end{equation}
Here, $\hbar$ is the reduced Planck constant and $e$ is the elementary charge. The best-fit values of the normal-state square resistance $R_{SN}$ and the BCS mean-field transition temperature $T_{C0}$ are listed in Table~\ref{tab:transport_SC_parameters}; the best-fit values of $A_{2d}$ are 2.21 and 2.16 for the film $\#$L135 and $\#$L134, respectively.

\begin{figure}[h!]
	\centerline{\includegraphics[width=0.5\textwidth]{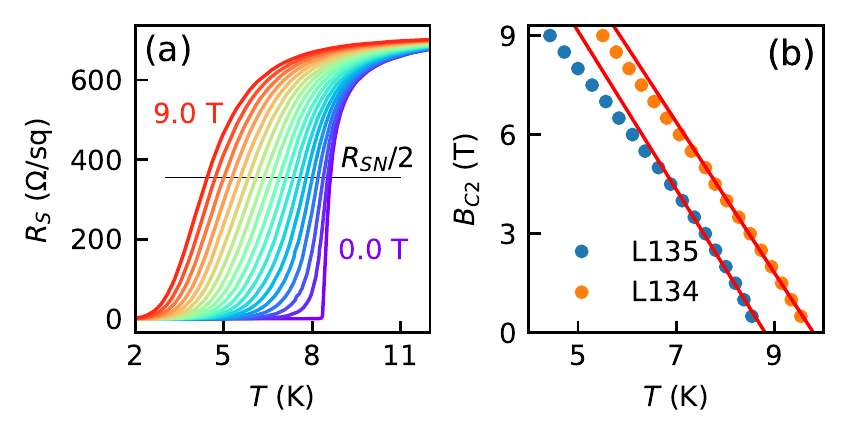}}
	\caption{(a) Square resistance for the film $\#$L135 vs temperature at a set of fixed magnetic fields in steps of 0.5~T. The straight horizontal line drawn at $R_S=R_{SN}/2$ defines the midpoint transition temperature for each field. (b)  Upper critical magnetic field as function of temperature. Points: experimental data; lines: the best-linear fits, which are used to evaluate $D$ and $B_{C2}(0)$. }
	\label{fig:Rs_BT}
\end{figure}

An external magnetic field suppresses the superconducting transition as it is shown in Fig.~\ref{fig:Rs_BT}(a) for a one representative sample $\#$L135. From these measurements, we determined the temperature-dependent upper critical field, $B_{C2}(T)$, shown in Fig.~\ref{fig:Rs_BT}(b), as a preset field at the midpoint-transition temperature $T_C$, i.e. at the temperature where $R_S = R_{SN}/2$. From the best-linear fits of the experimental $B_{C2}(T)$ data at temperatures close to $T_C$ ($T/T_C$ between 0.05 and 0.25), we extracted the slopes $\text{d} B_{C2} / \text{d} T$. We further used them to compute the electron diffusion coefficient as $D = 4 k_B/(\pi e) [\text{d}B_{C2}/\text{d}T]^{-1}$, the zero-temperature upper critical field as $B_{C2}(0) = -0.69 T_{C0} \text{d}B_{C2}/\text{d}T$ for the dirty regime, and the zero-temperature Ginzburg-Landau coherence length $\xi(0) = \sqrt{\Phi_0/(2 \pi B_{C2}(0))}$, where $k_B$ is the Boltzmann constant and $\Phi_0 = h/(2e)$ is the magnetic flux.  Total electron density  of  states  at  the  Fermi  energy $N(0)$  was determined via the Einstein relation $N(0) = 1/(D e^2 R_{SN} d)$ where $d$ is the film thickness. All these parameters are listed in Table~\ref{tab:transport_SC_parameters}.

\begin{figure}[h!]
	\centerline{\includegraphics[width=0.45\textwidth]{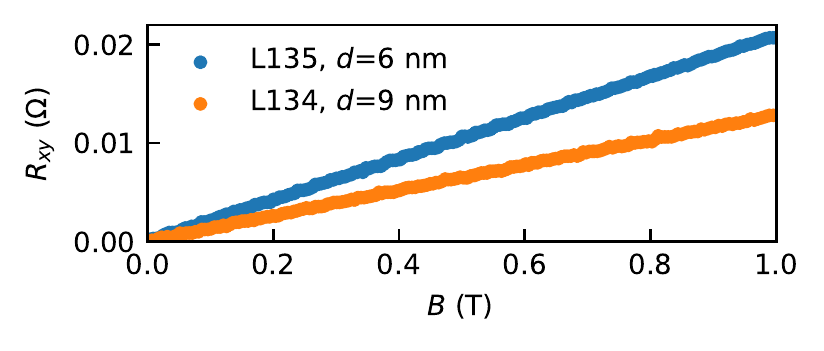}}
	\caption{Hall resistance vs magnetic field at 25~K. }
	\label{fig:R_xy}
\end{figure}

Magnetic field dependence of the Hall resistance $R_{xy}(B)$, i.e. the resistance appearing transverse to the current flow, is shown in Fig.~\ref{fig:R_xy}. It was measured at 25~K, well above the superconducting transition, where $R_{xy}$ changes linearly with $B$. From the slope of the $R_{xy}(B)$ dependence, we found the Hall coefficient as $R_H =  R_{xy}d / B$ that provides us the carrier density $n_e$ according to $R_H = 1 / (e n_e)$. Further, knowing the $n_e$, we determined the Fermi wavevector as $k_F = (3 \pi^2n_e)^{1 / 3}$, the elastic electron mean free path as $l_e = \hbar k_F/ (n_e e^2 R_{SN} d)$, and, finally, the Ioffe-Regel parameter $k_F l_e$. The latter indicates a slightly different level of disorder in our films. All these parameters are listed in Table~\ref{tab:transport_SC_parameters}.

\subsection{Magnetoconductance}

\begin{figure}[h!]
	\centerline{\includegraphics[width=0.5\textwidth]{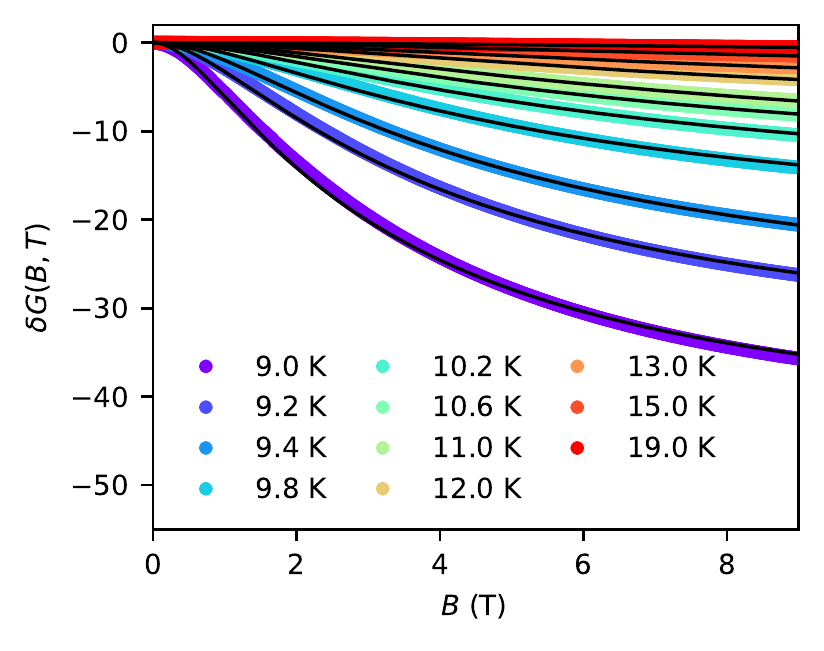}}
	\caption{Dimensionless magnetoconductance vs magnetic field for a set of fixed temperatures indicated in the legend for the film $\#$L135. Points: experimental data; black curves: the best fits with a sum of Eqs.~(\ref{eq:WL}-\ref{eq:DOS}).}
	\label{fig:MC_data}
\end{figure}

We determined the \textit{e-ph} scattering time for our NbTiN films by means of the magnetoconductance technique described in \cite{sidorova2020electron}. We measured $R_S$, varying the magnetic field from 0 to 9~T at a set of fixed temperatures from 9 to 20~K. From these data, we computed the dimensionless magnetoconductance as $\delta G(B,T)=2\pi^2\hbar/e^2\left(R_S^{-1}(B,T) - R_S^{-1}(0,T)\right)$. The result is shown in Fig.~\ref{fig:MC_data} with points for the film $\#$L135. The change in the conductance induced by the magnetic field originates from quantum interference effects and is described by the theory of quantum corrections to conductivity (see Appendix~\ref{app: QC}). We considered four quantum corrections given by Eqs.~(\ref{eq:WL}-\ref{eq:DOS}), which account for the electron weak-localization effect (WL), Aslamazov-Larkin (AL) and Maki-Thompson (MT) superconducting fluctuations, and electronic density of states (DOS) fluctuations. At experimental fields below $B_{C2}(0)$, the correction due to renormalization of the single-particle diffusion coefficient is small \cite{glatz2011quantum, minkov2004magnetoresistance} and we have neglected it.

\begin{figure}[h!]
	\centerline{\includegraphics[width=0.5\textwidth]{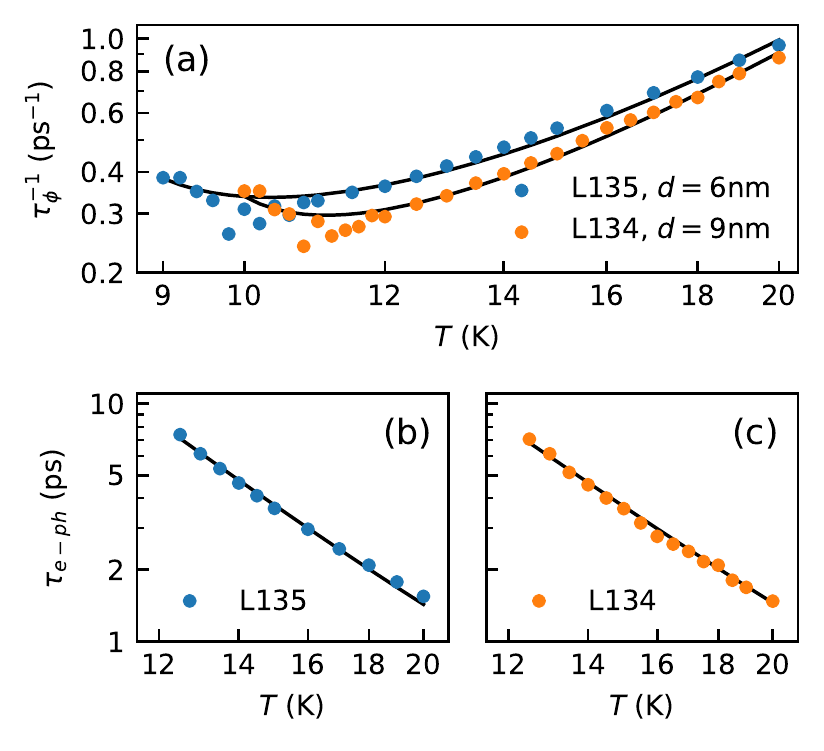}}
	\caption{(a) Electron dephasing rate vs temperature. Points: the best-fit values of $\tau_\phi^{-1}(T)$ obtained by fitting the experimental $\delta G(B,T)$ data; solid curves: the best fits with Eq.~(\ref{rates}). (b and c) Electron-phonon scattering time vs temperature for the NbTiN film $\#$L135 (b) and $\#$L134 (c) in a double-logarithmic scale. Points: values of $\tau_{e-ph}(T)$ obtained from the experimental $\tau_\phi^{-1}$ data as $\tau_{e-ph} = (\tau_\phi^{-1} - \tau_{e-e}^{-1} - \tau_{e-fl}^{-1})^{-1}$; solid lines: the best fits with the SM model, Eqs.~(\ref{eq:SM_tau_e-ph_l},\ref{eq:SM_tau_e-ph_t}).}
	\label{fig:rate_tau}
\end{figure}

We fitted the experimental $\delta G(B,T)$ data with a sum of Eqs.~(\ref{eq:WL}-\ref{eq:DOS}) using three independent fitting parameters, namely the temperature-dependent electron dephasing time $\tau_\phi(T)$, the temperature-independent  spin-orbit interaction time $\tau_{s.o.}$, and a constant $C^*$ allowing the latter to take values between 1.0 to 3.0 according to \cite{bergmann1984quantum, redi1977two, abrahams1971effect, tinkham1980introduction}. The best fits, shown with black curves in Fig.~\ref{fig:MC_data} for the representative film $\#$L135, were obtained with $C^* \approx 1.0$, $\tau_{s.o.}=20\pm5$~ps, and a maximum $\tau_\phi(T)$ of about 4~ps. For both films, $\tau_{s.o.}$ is of the order of the maximum $\tau_\phi$ that corresponds to the weak spin-orbit interaction in our NbTiN films. The best-fit values of $\tau_\phi^{-1}(T)$ are plotted in Fig.~\ref{fig:rate_tau}(a). 

The experimental dependence $\tau_\phi^{-1}(T)$ is described with a sum of rates affiliated with different phase-breaking scattering events in which electrons are involved. In the absence of extrinsic phase-breaking sources, these are electron-electron (\textit{e-e}) \cite{altshuler1982effects}, electron-phonon (\textit{e-ph}) \cite{sergeev2000electron}, and electron-fluctuation (\textit{e-fl}) \cite{brenig1985inelastic} scattering events. Consequently,
\begin{subequations}\label{rates} 
	\begin{equation}
	\tau_{\phi}^{-1} = \tau_{e-e}^{-1} + \tau_{e-ph}^{-1} + \tau_{e-fl}^{-1},
	\tag{\ref{rates}}
	\end{equation}
where
\begin{align}
	\tau_{e-e}^{-1} =	\frac{k_B T}{\hbar} \frac{1}{2C_1}   \ln(C_1), \label{rate_ee} \\
 	\tau_{e-ph}^{-1}=	\alpha_{e-ph}^{-1} \left(T/T_{C0} \right)^n, \label{rate_eph} \\
 	\tau_{e-fl}^{-1} =	\frac{k_B T}{\hbar}\frac{1}{2C_1} \frac{2\ln(2)}{\ln(T/T_{C0})+C_2}. \label{rate_efl}
 \end{align}
\end{subequations}
Here, $C_1=\pi \hbar / (R_{SN} e^2 )$ and $C_2 = 4 \ln(2)/[\sqrt{\ln(C_1)^2 + 128C_1/\pi} - \ln(C_1)]$. The expression for the \textit{e-e} dephasing rate, Eq.~(\ref{rate_ee}), accounts for dephasing only due to Nyquist noise, which dominates in our experimental temperature range $T\ll \hbar/(k_B \tau_e) \sim 10^4$~K ($\tau_e = l_e^2/(3 D)$ equals 0.3 and 0.5~fs for $\#$L135 and $\#$L135, respectively). It is worth noting that the expression for the \textit{e-ph} dephasing rate, Eq.~(\ref{rate_eph}), is valid in a relatively small temperature range, where $n$ is constant, in general, $n$ varies with the temperature. We fitted the experimental $\tau_\phi^{-1}(T)$ data shown in Fig.~\ref{fig:rate_tau}(a) with Eq.~(\ref{rates}) using two independent fitting parameters $\alpha_{e-ph}$ and $n$. At $T_{C0}$, $\tau_{e-ph}$ equals $\alpha_{e-ph}$. The best-fit values of $\alpha_{e-ph}$ and $n$ are listed in Table~\ref{tab:eph_parameters}, together with $\tau_{e-ph}$ values extrapolated to 10~K. 

\begin{table*}[ht!]
	\centering
	\caption{Some of the electron and phonon-related parameters of the studied here NbTiN films ($\#$L134 and $\#$L135) and films from two other materials NbN and WSi, for comparison. The values of $u_t$ and $\rho$ are obtained from fits with the SM model. The values of $\tau_{esc}$ and $c_e/c_{ph}$ are obtained from fits with the 2T model.}
	\label{tab:eph_parameters}
	\begin{tabularx}{\textwidth}{@{}cccccc|cc@{}}
		\hline \hline
		Material&Sample	&$\alpha_{e-ph}$\footnotemark & $n$\footnotemark    & $\tau_{e-ph}$(10~K)\footnotemark       & $\tau_{EP}(T_{C0})$\footnotemark          & $\tau_{esc}$  & $c_e/c_{ph}(T_{C0})$ \\
		&& (ps)  &  & (ps)  & (ps)                &   (ps)                 &             \\ 
		\hline
		NbTiN&L135      												&31.2 $\pm$ 3.5& 3.5 $\pm$ 0.1  & 16.9  & 4.9  &52.6 $\pm$ 1.5   &0.39 $\pm$ 0.04     \\
		NbTiN&L134												        &17.7 $\pm$ 1.1 & 3.4 $\pm$ 0.1  & 15.0  & 2.9  & 79.6 $\pm$ 4.3 & 0.13 $\pm$ 0.01    \\
		NbN\cite{sidorova2020electron}&2559      &9.3                    & 3.5                      & 11.9  & 1.4            	     & 25.9                      & 0.83 $\pm$ 0.18   \\
		NbN\cite{sidorova2020electron}&A853      & 21.7                    & 3.2                 & 12.4  & 4.2           	     & 39.0                      &0.25 $\pm$ 0.03   \\
		WSi\cite{zhang2016characteristics}&4     & 66                        & 3.0                    & 4.5   & 14.9                  & -                             &1.40 $\pm$ 0.30   \\
		\hline \hline
	\end{tabularx}
	\footnotetext[1]{At $T=T_{C0}$,  $\tau_{e-ph}$ equals $\alpha_{e-ph}$.}
	\footnotetext[2]{The exponent in the temperature dependence of $\tau_{e-ph}^{-1}\propto T^n$}
	\footnotetext[3]{The dephasing time due to \textit{e-ph} scattering (identical with the single-particle \textit{e-ph} scattering time \cite{rammer1986destruction}) extrapolated to 10~K.}
	\footnotetext[4]{The \textit{e-ph} energy relaxation time (proportional to the $\tau_{e-ph}$ with a proportionality coeff. computed with Eq.~(11) in \cite{il1998interrelation}).}
\end{table*}

Fig.~\ref{fig:rate_tau}(b,c) shows $\tau_{e-ph}$ obtained from the dephasing rates as $\tau_{e-ph} = (\tau_\phi^{-1} - \tau_{e-e}^{-1} - \tau_{e-fl}^{-1})^{-1}$ in the temperature range where $\tau_{e-ph}^{-1}$ dominates other dephasing rates. In the dirty limit with respect to \textit{e-ph} scattering characterized by the product $q_T l_e \ll 1$, where $q_T=k_B T/ \hbar u$ is the phonon wave vector and $u$ is the sound velocity, \textit{e-ph} scattering is described by a model developed by Sergeev and Mitin in \cite{sergeev2000electron} (the SM model, Appendix~\ref{app: SM}). The model provides the single-particle \textit{e-ph} scattering time, which is identical with $\tau_{e-ph}$ \cite{rammer1986destruction} extracted by the magnetoconductance method. We fitted experimental $\tau_{e-ph}$ data wit the SM model, Eq.~(\ref{eq:SM}), using two independent fitting parameters: the sound velocity of transverse phonons $u_t$ and the mass density $\rho$. We fixed other parameters such as $k_F$, $l_e$, and $N(0)$, at their values given in Table.~\ref{tab:transport_SC_parameters} and adopted for $m_e$ the value of the free electron mass. We implemented the lattice parameter $a_0=0.43$~nm \cite{makise2010characterization, arockiasamy2016ductility, hazra2018superconducting}, the sound velocity of longitudinal phonons $u_l = 2 u_t$ that is approximately valid for a large variety of materials, and
$k \sim 1.0$ (Appendix~\ref{app: SM}). The best-fit values of $u_t$ and $\rho$ are listed in Table~\ref{tab:phonon_properties}.

\subsection{Photoresponse in the time domain}

\begin{figure}[h!]
	\centerline{\includegraphics[width=0.5\textwidth]{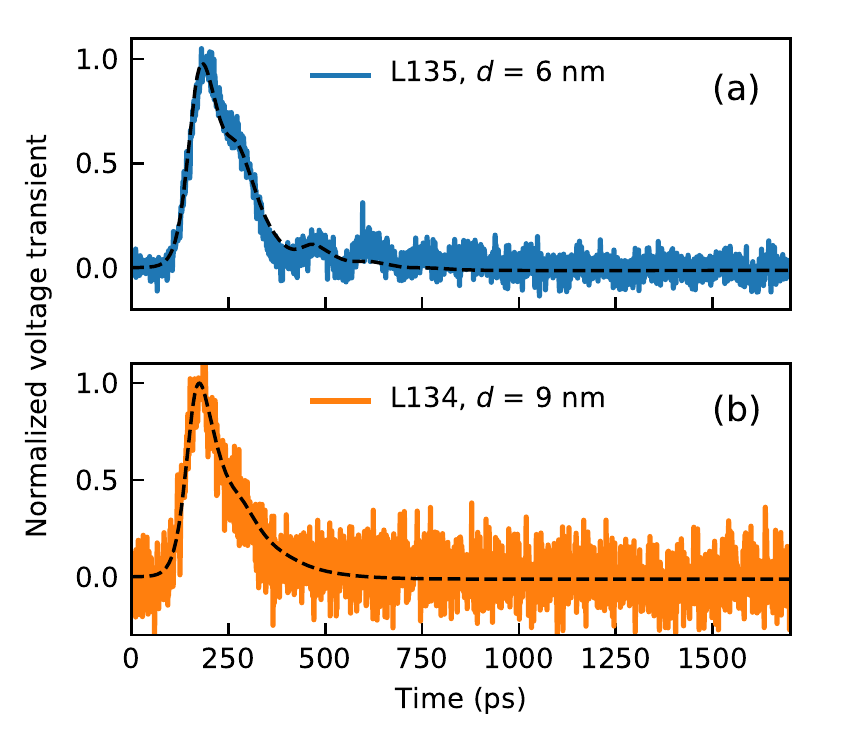}}
	\caption{Normalized voltage transient vs time for microbridges fabriacated from NbTiN film $\#$L135 (a) and $\#$L134 (b). Solid curves: experimental data; dashed curves: the best fits with the 2-T model, Eqs.~(9-13) in \cite{sidorova2020electron}.}
	\label{fig:photoresp}
\end{figure}

We studied the energy relaxation of nonequilibrium electrons in NbTiN microbridges by means of the photoresponse technique. The microbridges were fabricated from the two NbTiN films, $\#$L135 and $\#$L134, shaped to a width of $\sim$90~$\mu$m and a length of $\sim$1~$\mu$m (along the current flow) with tapered contacts toward the electrodes, which are 21~$\mu$m apart, to reduce the current-crowding effect. The microbridges were installed in a continuous flow cryostat, kept in the resistive state at an ambient temperature slightly larger than $T_{C0}$, and exposed to light pulses via a quartz window of the cryostat. Light was focused onto the microbridges into a spot with a diameter larger than $\sim100~\mu$m. The light pulses were generated by a Ti:Sapphire laser at a wavelength of about 800~nm with a repetition rate of 80~MHz and a subpicosecond duration. Biasing the microbridges with a small direct current, we recorded their photoresponse (amplified voltage transients) to light pulses in the time domain with a sampling oscilloscope (Fig.~\ref{fig:photoresp}). The overall bandwidth of our readout was 0.1~-~5~GHz. 

We described the voltage transients, shown in Fig.~\ref{fig:photoresp}, with the two-temperature (2T) model using the formalism given by Eqs.~(9-13) in \cite{sidorova2020electron}. This formalism takes into account the effect of a finite readout bandwidth and the signal ringing caused by impedance mismatch. The later is seen in Fig.~\ref{fig:photoresp}(a). The 2T model describes the evolution of electron and phonon effective temperatures raised by excitation via two coupled time-dependent equations with three independent parameters: the ratio between electron and phonon heat capacities $c_e/c_{ph}$, the phonon escape time $\tau_{esc}$, and the \textit{e-ph} energy relaxation time $\tau_{EP}$. We note here that $\tau_{EP}$ differs from $\tau_{e-ph}$ (both are in Table~\ref{tab:eph_parameters}) by a coefficient $\tau_{EP}/\tau_{e-ph}<1$, given by Eq.~(11) in \cite{il1998interrelation}, which depends on the exponent $n$, e.g., is about 0.6 for $n=2.0$ and 0.1 for $n=4.0$. Hence, describing the photoresponse with the 2T model, one gets indirect \textit{calorimetric} information on heat capacities.

We fitted the experimental transients using $c_e/c_{ph}$ and $\tau_{esc}$ as fitting parameters, along with the fixed values of $\tau_{EP}$. The results are shown in Fig.~\ref{fig:photoresp}. The best-fit values of $c_e/c_{ph}$ and $\tau_{esc}$, together with computed values of $\tau_{EP}$, are listed in Table~\ref{tab:eph_parameters}.

\section{Discussion}
\label{sec: Discussion}

For the studied thin NbTiN films, we found that in the temperature range from 12 to 20~K the \textit{e-ph} scattering time varies as $\tau_{e-ph}\propto T^{-n}$  with $n$ between 3.4 and 3.5 and that at 10~K the extrapolated value,  $\tau_{e-ph}$(10~K), falls between 15 and 17~ps. NbTiN is a compound of two well studied nitrides, NbN and TiN, which exhibit rather different scattering times. The $\tau_{e-ph}\propto T^{-3.0}$ was found for thin TiN films in the temperature range from 1.7 to 4.2~K \cite{kardakova2013electron} resulting in $\tau_{e-ph}$(10~K)~$\sim 1.8$~ns (actually, the authors derived the \textit{e-ph} energy relaxation time, i.e.,  $\tau_{EP}$ in our notation, which, along with the exponent $n$, we used to estimate $\tau_{e-ph}$). For NbN films with thicknesses from 5 to 10 nm, $n$ spreading between 3.2 and 3.8 was found  for the temperature range from 14 to 30~K along with the extrapolated value $\tau_{e-ph}$(10~K) in the range from 12 to 18~ps. \cite{sidorova2020electron}. Consequently, with respect to inelastic \textit{e-ph} interaction, our Nb$_x$Ti$_{1-x}$N films with the Nb/Ti fraction $x\approx0.6$ are closer to NbN than to TiN. This certainly indicates that the strength of \textit{e-ph} coupling in Nb$_x$Ti$_{1-x}$N films cannot be controlled by a fine change of $x$ in the range from 0.6 to 1.0.

Let us now discuss the $\tau_{e-ph}$ provided by the SM model (Appendix~\ref{app: SM}). Under the condition $q_T l_e \ll 1$ and the dominance of vibrating scattering centers ($k\sim1$) fulfilled for our films, $\tau_{e-ph(l)} \gg \tau_{e-ph(t)}$ and, consequently, the total \textit{e-ph} scattering rate is controlled by the interaction of electrons with transverse phonons $\tau_{e-ph} \equiv \tau_{e-ph(t)}$. Therefore, we limit the discussion to the properties of transverse phonons. Interacting with a phonon, an electron exchanges momentum $\sim q_T$ within an interaction region $\sim 1/q_T \propto u_t$, so that the smaller the phonon velocity, the smaller the interaction region, and the shorter the interaction time. In fact, $\tau_{e-ph(t)}\propto \rho u_t^3$ (Eq.~\ref{eq:SM_tau_e-ph_t}). With $\rho^*=7.5$~g/cm$^3$ and $u_t^*=4.7$~nm/ps for bulk crystalline NbTiN computed from first-principles \cite{arockiasamy2016ductility} (Table~\ref{tab:phonon_properties}), the SM model provides two orders of magnitude larger $\tau_{e-ph}$ than the experimental value obtained from magnetoconductance (Table~\ref{tab:eph_parameters}). The best-fit values of $\rho$ and $u_t$, which provide experimetal values of $\tau_{e-ph}$ in the framework of the SM model are listed in Table~\ref{tab:phonon_properties}. The same procedure we applied to fit the $\tau_{e-ph}$ data obtained for thin WSi films by magnetoconductance measurements in \cite{zhang2016characteristics} (see Appendix~\ref{app: WSi}). The best-fit values of $u_t$ and $\rho$ for WSi are also listed in Table~\ref{tab:phonon_properties}.

\begin{table*}[t!]
	\centering
	\caption{Phonon-related properties. $c_{ph}^{(*)}$ were computed as for 3d Debye phonons using corresponding sound velocities $u_t^{(*)}$. $c_{ph}^{exp}$ were obtained from $c_e/c_{ph}$ ratios found as the best fits of the 2T model to the photoresponse data taking $c_e$ predicted by the Drude model.}
	\label{tab:phonon_properties}
	\begin{tabularx}{\textwidth}{cc|cccc|ccc|c}
		\hline\hline
		Material  & Sample & $u_t^*$\footnotemark[1]                   & \multicolumn{1}{c}{$u_l^*$\footnotemark[1] } & $\rho^*$\footnotemark[1]                  & $c_{ph}^*$            & $u_t$\footnotemark[2]   & $\rho$\footnotemark[2]    & $c_{ph}$              & $c_{ph}^{exp}$        \\
		&        & (nm/ps)                  & \multicolumn{1}{c}{(nm/ps)} & (g/cm$^3$)               & (J/Km$^{3}$) & (nm/ps) & (g/cm$^3$) & (J/Km$^{3}$) & (J/Km$^{3}$) \\
		\hline
		NbTiN & L135   & \multirow{2}{*}{5.3-4.7} & \multirow{2}{*}{9.0-8.4}    & \multirow{2}{*}{6.7-7.5} &    400-500                   & 1.8     & 3.0       &           8800            &           2700            \\
		NbTiN & L134   &                          &                             &                          &         500-700              & 1.9     & 3.6        &         10800             &           11000            \\
		NbN\cite{sidorova2020electron}   & 2259   & \multirow{2}{*}{4.4}     & \multirow{2}{*}{8.0}        & \multirow{2}{*}{8.2}     &       1300                & 2.4     & 7.8        &          7800             &              2500        \\
		NbN\cite{sidorova2020electron}   & A853  &                          &                             &                          &         600              & 2.2     & 5.2        &        4700               &          4000             \\
		WSi\cite{zhang2016characteristics}   &    4    & 3.0                  & \multicolumn{1}{c}{5.4} & 15.8              &          200             & 2.1\footnotemark[3]     & 8.2\footnotemark[3]        &          630             &       560          \\  
		\hline\hline   
	\end{tabularx}
	\footnotetext[1]{Computed from first principles for NbTiN \cite{arockiasamy2016ductility}, NbN \cite{arockiasamy2016ductility}, and W$_3$Si \cite{pan2017probing} with crystal structures.}
	\footnotetext[2]{Best-fit values of the parameters in the SM model.}
	\footnotetext[3]{Obtained in Appendix~\ref{app: WSi}.}
\end{table*}

Let us now estimate the phonon escape time using the acoustic mismatch model \cite{kaplan1979acoustic}. First, we compute the transmission $\bar{\eta} \approx 0.2$ of the film/substrate interface for phonons using the best-fit values of $u_t$ and $\rho$ (Table~\ref{tab:phonon_properties} for our NbTiN films; for SiO$_2$ substrate, we took them from Table~I in \cite{kaplan1979acoustic}). Further, we find $\tau_{esc} = 4 d / (\bar{\eta} \bar{u}) \approx 60.6$~ps for the film $\#$L135 and 81.9~ps for the film $\#$L134, where $\bar{u} \approx u_t$ is the weighted sound velocity. Hence, in NbTiN films the phonon escape time scales with the film thickness as $\tau_{esc}$[ps]~$\approx 9\,d$[nm]. The computed values of $\tau_{esc}$ are in good agreement with those found as best fits to the photoresponse data in the framework of the 2T model (Table~\ref{tab:eph_parameters}). This allows us to conclude that the SM model provides reasonable values of $u_t$ and $\rho$.

Further, from the best-fit values of $c_e/c_{ph}$ ratios (Table~\ref{tab:eph_parameters}) and $c_e=\pi^2 k_B^2 N(0) T/3$ predicted by the Drude model, we find the phonon heat capacities $c_{ph}^{exp}$, which are listed in Table~\ref{tab:phonon_properties}. Here we assumed that the Drude model predicts a correct value of $c_e$, relying on the fact that our NbTiN films are three-dimensional (3d) with respect to electron transport ($d \gg l_e$).

It is clearly seen from Table~\ref{tab:phonon_properties} that the best-fit values of $u_t$ for thin NbTiN, NbN, and WSi films are systematically reduced as compared to those of corresponding bulk crystalline materials (marked with the asterisk in Table~\ref{tab:phonon_properties}) by $(u_t^*-u_t)/u_t^* \sim40-60\%$. Moreover, for all these films, both experimental  (calorimetric) $c_{ph}^{exp}$ and $c_{ph}$ computed in the framework of the Debye model for 3d phonons with the reduced sound velocities as $c_{ph} = 2/5\,\pi^2 k_B (k_B T/ \hbar)^3 u_{av}^{-3}$, where $u_{av} = [2/3 u_t^{-3}+ 1/3 u_l^{-3}]^{-1/3} \approx u_t$, are systematically larger than the heat capacities $c_{ph}^*$ computed within the same approximation with the sound velocities of corresponding bulk crystalline materials.


There are three effects on the sound velocity and phonon heat capacity to be discussed: (i) depletion of long-wavelength phonon states,(ii) phonon softening, and (iii) film morphology (amorphous or granular structure). 

	At low temperatures, the wavelength of a thermal phonon $\lambda_{ph} \approx 2\pi\hbar u/(k_B T)$ becomes comparable or larger than the film thickness that modifies the 3d Debye phonon spectrum. Phonon spectra in a metallic film with a variable thickness on a semi-infinite substrate were computed and compared with experimental data \cite{frick1975phonon}. For phonons excited perpendicularly to the film/substrate interface, by lowering the film thickness at low temperatures, the authors found a strong depletion of long-wavelength (low-energy) phonon states beyond the cut-off wavelength $\lambda_{ph}=2d$. Qualitatively the depletion should not affect the sound velocity but only cause a reduction of the phonon heat capacity; its impact should grow with the increase in the ratio $\lambda_{ph}/d$. A similar but even stronger impact of the depletion on the phonon heat capacity should occur in granular films where the depletion is almost isotropic. It affects phonon states with all directions of the wave vector. Even though for all the films in Table~\ref{tab:phonon_properties} the ratio $\lambda_{ph}/d>1$, the depletion alone does not fully explain our experimental results. 
		
		In a thin mono-crystalline film, phonon softening, i.e. the decrease of the effective Debye temperature with respect to the bulk value, occurs due to weakening of ion bounds at film surfaces \cite{lang2005finite}. Grain boundaries and defects increase the relative amount of weak bounds and enhance phonon softening. \cite{ma2010electrical, cheng2015temperature}. Clearly, the weakening of bounds leads to a reduction in the mean sound velocity and a corresponding increase in the phonon heat capacity. The effect is most pronounced in ultra-thin granular films due to a large surface-to-volume ratio. We have to note here that there exists an alternative explanation of this effect relying on the proximity effect between the film and the oxide layers on the film surfaces \cite{engel2006electric}.
		
		As it is seen from the discussion above, the effect of the film morphology on the sound velocity and phonon heat capacity is hard to separate from the effect of the reduced dimensionality. The structure of both NbTiN films \cite{zichi2019optimizing} and NbN films \cite{engel2006electric, ilin2008ultra, lin2013characterization} is polycrystalline, granular with the mean grain size of the order of the film thickness or larger. Polycrystalline NbTiN  and NbN films are composites of crystalline grains with amorphous boundaries \cite{zhang2009grain} while WSi films are rather amorphous (verified by x-ray diffraction in 60~nm-thick WSi films \cite{baek2011superconducting}) without pronounced grains. However, it has been recently shown that 6~nm-thick WSi films exhibit a preferred orientation rather than an amorphous state \cite{jin2019preparation}.  For longitudinal phonons, the amorphous phase is not expected to reduce noticeably the sound velocity, while for transverse phonons it does because of substantial weakening of shear modulus. A large reduction of transverse sound velocities in amorphous materials, relative to the crystalline materials, was observed, for instance, in \cite{golding1972soft}. This effect can be considered as amorphous phonon softening. It is known \cite{bergmann1976amorphous} that for amorphous materials it results in an increase in the phonon heat capacity just by the factor ($c_{ph} \propto u_t^{-3}$) the sound velocity is reduced in an amorphous state as compared to crystalline material.
		
		Comparing calorimetric experimental values of $c_{ph}^{exp}$ with the $c_{ph}$ values computed with reduced sound velocities $u_t$ of the SM model (Table~\ref{tab:phonon_properties}), we found that for NbN ($\#$A853),  
		WSi, and NbTiN ($\#$L134) films $c_{ph}^{exp} \approx c_{ph}$ that might be an indirect indication of their amorphous state. Although the statement may sound conceivable for the WSi film and for the NbN, $\#$A853 film (see the discussion on this particular film in \cite{sidorova2020electron}),  it contradicts to the known polycrystalline structure of the NbTiN, $\#$L134 film. For other granular films (NbTiN, $\#$L135 and NbN, $\#$A-2259), $c_{ph}^{exp}$ is noticeably smaller than $c_{ph}$. The most plausible explanation of this difference is the depletion of long-wavelength phonon modes in granules. We presume that like for NbN films grown on non-heated substrates \cite{engel2006electric}, the grain size in NbTiN films increases with the increase of the film thickness. Hence, in the thicker NbTiN film ($\#$L134, $d$=9~nm), the phonon depletion is less pronounced than in the thinner film ($\#$L135, $d$=6~nm) and the phonon heat capacity in the former is defined by the phonon softening due to film surfaces. We suppose that in the ultra-thin polycrystalline films analyzed here, the impact of the film morphology on the phonon heat capacity depends on the grain size and may noticeably modify the impact of film dimensionality. Another important observation is that the mass densities of all thin films reported here are noticeably less than those of corresponding crystalline materials. This is the direct consequence of the film morphology since in an amorphous or polycrystalline film host atoms are packed at a smaller density than in the corresponding crystalline material.

Finally, let us compare the properties of our NbTiN films and NbN films studied in \cite{sidorova2020electron} (samples $\#$2259 and $\#$A853) that are relevant to SNSPD and HEB practical devices. With respect to superconducting and transport properties, together with \textit{e-ph} scattering times, both materials are very similar. The main difference is in their phonon escape times, which, in NbTiN/SiO$_2$ (this study) $\tau_{esc}$[ps]~$\approx 9\,d$[nm] is about twice  as large as in NbN/SiO$_2$ ($\tau_{esc}$[ps]~$\approx 5\,d$[nm]) \cite{sidorova2020electron}. Besides the timing jitter, the difference in $\tau_{esc}$ is not expected to somewhat noticeably affect the performance of an SNSPD. However, this difference severely affects the performance of an HEB. Specifically, its intermediate-frequency (IF) bandwidth is controlled by the cooling rate of electrons, in which the phonon escape rate plays a limiting role \cite{semenov2002hot, klapwijk2017engineering}. A larger $\tau_{esc}$ results in a smaller IF bandwidth. This explains why, despite all the efforts, NbTiN-based HEBs repeatedly demonstrate smaller IF bandwidth than NbN-based HEBs (\cite{shurakov2015superconducting}).

\section{Conclusion}

We have studied dephasing and energy relaxation of electrons in thin superconducting NbTiN films on Si/SiO$_2$ substrates with magnetoconductance and photoresponse techniques, respectively, and compared our results with those reported in the literature for NbN and WSi films. Our main findings are the following:

(a) The studied NbTiN films are strongly disordered with respect to electron-phonon scattering, the product $q_T l_e (T_C) \ll 1$, and electron transport, the Ioffe-Regel parameter $k_F l_e = 2.5-3.0$. In the temperature range from 12 to 20~K, their inelastic \textit{e-ph} scattering rate varies as $\tau_{e-ph}^{-1}\propto T^n$ with $n=3.4-3.5$ and amounts (extrapolated) to 15.0-16.9~ps at 10~K. The phonon escape time varies with the film thickness as $\tau_{esc}$[ps]~$\approx 9\,d$[nm]. 

(b) In all studied and considered films, we have found a systematic reduction of the sound velocity by $\Delta u/u=40-60$~\% as compared to the sound velocities computed from the first principles for corresponding bulk crystalline materials.

(c) For all films the experimental calorimetric heat capacities of phonons are much larger than the heat capacities computed for corresponding crystalline materials with the 3d Debye model. For amorphous films and relatively thick polycrystalline films, calorimetric heat capacities are almost equal to those computed with the reduced sound velocities, while for granular films they are a few times less. The latter is most plausibly due to the depletion of low energy phonon modes in granules.

\section*{Acknowledgements:}
Authors greatly acknowledge the support of V. Zwiller in the sample preparation and stimulating discussions with K. Illin  on the magnetic properties and morphology of superconducting granular films.

\appendix

\section{Quantum corrections}
\label{app: QC}
The theory of quantum corrections to the classical Drude conductance is applicable for materials with $k_F l_e > 1$. 
With respect to the characteristic scales of the theory, i.e. the thermal coherence length $L_T=\sqrt{2\pi \hbar D/(k_B T)}$ and the electron dephasing length $L_\phi$, our NbTiN films belong to quasi-2d systems ($d < L_T, L_\phi$). Therefore, the analytical expressions we use to compute theoretical (dimensionless) corrections correspond to the 2d limit. For weak spin-orbit interaction and in the absence of magnetic scattering, the WL correction \cite{hikami1980spin, bergmann1984quantum} is given by
\begin{equation}\label{eq:WL}
		\delta G^{WL}(B,T) = \frac{3}{2} Y \left( \frac{B}{\frac{4}{3}B_{s.o.} + B_{\phi}} \right) - \frac{1}{2} Y\left(\frac{B}{B_{\phi}}\right).
\end{equation}
Here, the function $Y(x) = \ln(1/x)+\psi(1/2 + 1/x)$ is defined via the digamma function $\psi(x)$. The characteristic fields are defined as $B_j = \hbar/(4 e D \tau_j)$, where the indices 's.o.' and '$\phi$' stand for spin-orbit and dephasing scattering, respectively. The AL correction \cite{aslamasov1968influence} is given by 
\begin{dmath}\label{eq:AL}
		\delta G^{AL}(B,T)=\frac{\pi^2}{2 \ln(T/T_{C0})}
		\left\{
		\frac{B_C}{B} \left( 1 - 2\frac{B_C}{B}  \\ \left[\psi \left(1 + \frac{B_C}{B} \right) - 
		\psi \left(\frac{1}{2} + \frac{B_C}{B} \right) \right] \right) - \frac{1}{4}
		\right\}.
\end{dmath}
Here, $B_C=C^*\hbar/(4eD\tau_{GL})$, where the Ginzburg-Landau time $\tau_{GL}=(\pi\hbar)/(8k_B T \ln(T/T_{C0}))$ represents the lifetime of Cooper pairs. The MT correction \cite{maki1968critical,thompson1970microwave, larkin1980reluctance, dos1985superconducting} is given by 
\begin{equation}\label{eq:MT}
		\delta G^{MT}(B,T) = -\beta_{LSA}(T) \left[ Y\left(\frac{B}{B_{\phi}}\right) - Y\left(\frac{B}{B_C}\right)\right],
\end{equation}
where the parameter $\beta_{LSA}(T)=2\pi k_B T\hbar^{-1} (1/\tau_{GL} -1/\tau_{\phi})^{-1}$. Finally, the DOS correction \cite{glatz2011quantum, larkin2003fluctuation} is given by
\begin{equation}\label{eq:DOS}
		\delta G^{DOS}(B,T) = \frac{28\varsigma(3)}{\pi^2} Y\left(\frac{B}{B_C}\right),
\end{equation}
where $\varsigma(3)=1.202$ is the Riemann zeta function. The total theoretical magnetoconductance, i.e. the sum of the four terms in Eqs.~(\ref{eq:WL}-\ref{eq:DOS}), is used to describe the experimental $\delta G(B,T)$ data.

\section{Inelastic \textit{e-ph} (single-particle) scattering time}
\label{app: SM}
The degree of disorder with respect to \textit{e-ph} scattering is characterized by the product $q_T l_e$. For our NbTiN films, in the experimental temperature range, $q_T l_e \ll 1$ that corresponds to the strong disordered regime. The SM model \cite{sergeev2000electron} describes \textit{e-ph} scattering in disordered metals and provides the inelastic (single-particle) scattering rate of electrons at the Fermi level via interaction with phonons of different polarizations:
\begin{subequations}\label{eq:SM} 
	\begin{equation}
	\tau_{e-ph}^{-1} = \tau_{e-ph(l)}^{-1} + \tau_{e-ph(l)}^{-1},
	\tag{\ref{eq:SM} }
	\end{equation}
	where interaction with \textit{longitudinal} phonons is accounted in
	\begin{equation}
	\tau_{e-ph(l)}^{-1} = \frac{7 \pi \zeta(3)}{2 \hbar} \frac{\beta_l(k_B T)^3}{(p_F u_l)^2}
	F_l (q_{T(l)} l_e),		\label{eq:SM_tau_e-ph_l}
	\end{equation}
	and with \textit{transverse} phonons (two polarizations are taken into account) in
	\begin{equation}
	\tau_{e-ph(t)}^{-1} = 
	3 \pi^2 \frac{\beta_t(k_B T)^2}{(p_F u_t)(p_F l_e)} k F_t (q_{T(t)} l_e). \label{eq:SM_tau_e-ph_t}
	\end{equation}
\end{subequations}
The indices '$l$' and '$t$' denote values associated with \textit{longitudinal} and \textit{transverse} phonon modes. Here, $\beta_{l(t)}=(2E_F /3 )^2 (N(0)/(2\rho u_{l(t)}^2))$ is the dimensionless coupling constant, $E_F=(p_F)^2/(2 m_e)$ is the Fermi energy, $p_F = \hbar k_F$ is the Fermi momentum, and $m_e$ is the electron mass. In Eq.~\ref{eq:SM_tau_e-ph_l}, the integral $F_l(z) = \frac{2}{7 \zeta(3)} \int_{0}^{A_l} dx \,\Phi_l(xz) [N(x) + n(x)] x^2$, where $N(x)$ and $n(x)$ are Bose and Fermi distribution functions, and $\Phi_l (x) = \frac{2}{\pi} \left( \frac{x \arctan(x)}{x - \arctan(x)} -  \frac{3}{x} k \right)$ is the Pippard function. In Eq.~\ref{eq:SM_tau_e-ph_t}, the integral $F_t(z) = \frac{4}{\pi^2} \int_{0}^{A_t} dx \, \Phi_t(xz) [N(x) + n(x)] x$, where $\Phi_t (x) = 1 + k [3x -3(x^2 +1)\arctan(x)]/[2 x^3]$.  The upper limit of the integrals $F(z)$ is $A_{l(t)}=(6 \pi^2 )^{1/3} (a_0\, q_{T,l(t)})^{-1}$. The parameter $ 1 \geq k \geq 0$ reveals the property of electron scattering centers ($k = 1$ corresponds to scattering centers vibrating in the same way as the host lattice, e.g. light impurities, $k = 0$ to the static scattering centers, e.g. heavy impurities and rigid boundaries). 

The total inelastic \textit{e-ph} scattering rate given by Eq.~(\ref{eq:SM}) is used to describe the experimental $\tau_{e-ph}$ data shown in Fig.~\ref{fig:rate_tau}(a,b).

\section{WSi films}
\label{app: WSi}
We describe the $\tau_{e-ph}(T)$ data obtained by the magnetoconductance method for WSi films in \cite{zhang2016characteristics} using the SM model (Appendix~\ref{app: SM}). Following the notation here, the authors of \cite{zhang2016characteristics} found $\alpha_{e-ph} \equiv \tau_{e-ph}(T_{C0}) = 66$~ps and $n=3$ between 5 and 20~K which we added to Table~\ref{tab:eph_parameters}. Fitting the SM model to these data, we used two  independent  parameters $u_t$ and $\rho$; other parameters as $D$ and $N(0)$ were fixed at their values provided in \cite{zhang2016characteristics} (also listed in Table.~\ref{tab:transport_SC_parameters}). We used $a_0=0.46$~nm \cite{mattheiss1992calculated},  $k_F=N(0) \pi^2 \hbar^2/m_e$ computed wit the free electron mass $m_e$, $l_e \sim 0.2$~nm, and $k\sim1.0$. The best-fit values of $u_t$ and $\rho$ are listed in Table~\ref{tab:phonon_properties}. 

\bibliography{sample}

\end{document}